\documentclass[sigconf]{acmart}
\def\BibTeX{{\rm B\kern-.05em{\sc i\kern-.025em b}\kern-.08emT\kern-.1667em\lower.7ex\hbox{E}\kern-.125emX}}
    
\usepackage{nicefrac}
\usepackage{siunitx}
\usepackage{array,framed}
\usepackage{booktabs}
\usepackage{
  color,
  float,
  epsfig,
  wrapfig,
  graphics,
  graphicx,
  subcaption,
}
\usepackage{textcomp}
\usepackage{setspace}
\usepackage{latexsym,fancyhdr,url}
\usepackage{enumerate}
\usepackage{algorithm2e}
\usepackage{algpseudocode}
\usepackage{graphics}
\usepackage{xparse} % argument parsing -- \edist
\usepackage{xspace}
\usepackage{multirow}
\usepackage{csvsimple}
\usepackage{balance}
\usepackage[utf8]{inputenc}
\usepackage{
  tikz,
  pgfplots,
  pgfplotstable
}
\usepackage{hyperref}

\usetikzlibrary{
  shapes.geometric,
  arrows,
  external,
  pgfplots.groupplots,
  matrix
}

\pgfplotsset{compat=1.9}
\usepackage{mathtools,}

\DeclareMathAlphabet{\mathcal}{OMS}{cmsy}{m}{n}
\DeclareGraphicsExtensions{%
    .png,.PNG,%
    .pdf,.PDF,%
    .jpg,.mps,.jpeg,.jbig2,.jb2,.JPG,.JPEG,.JBIG2,.JB2}

\usepackage{xparse}
\newcommand{\bnm}{\begin{newmath}}
\newcommand{\enm}{\end{newmath}}

\newcommand{\bea}{\begin{eqnarray*}}%
\newcommand{\eea}{\end{eqnarray*}}%

\newcommand{\bne}{\begin{newequation}}
\newcommand{\ene}{\end{newequation}}

\newcommand{\bal}{\begin{newalign}}
\newcommand{\eal}{\end{newalign}}

\newenvironment{newalign}{\begin{align}%
\setlength{\abovedisplayskip}{4pt}%
\setlength{\belowdisplayskip}{4pt}%
\setlength{\abovedisplayshortskip}{6pt}%
\setlength{\belowdisplayshortskip}{6pt} }{\end{align}}

\newenvironment{newmath}{\begin{displaymath}%
\setlength{\abovedisplayskip}{4pt}%
\setlength{\belowdisplayskip}{4pt}%
\setlength{\abovedisplayshortskip}{6pt}%
\setlength{\belowdisplayshortskip}{6pt} }{\end{displaymath}}

\newenvironment{newequation}{\begin{equation}%
\setlength{\abovedisplayskip}{4pt}%
\setlength{\belowdisplayskip}{4pt}%
\setlength{\abovedisplayshortskip}{6pt}%
\setlength{\belowdisplayshortskip}{6pt} }{\end{equation}}

\newcounter{ctr}

\newcounter{mytable}
\def\mytable{\begin{centering}\refstepcounter{mytable}}
\def\endmytable{\end{centering}}

\newcounter{myfig}
\def\myfig{\begin{centering}\refstepcounter{myfig}}
\def\endmyfig{\end{centering}}

\newlength{\saveparindent}
\newlength{\saveparskip}
\newcommand{\E}{{\rm I\kern-.3em E}}

\renewcommand{\eqref}[1]{\mbox{Equation~(\ref{#1})}}

\def \part {part}

\renewcommand{\paragraph}[1]{\vspace*{6pt}\noindent\textbf{#1}\;}

\def \blackslug{\hbox{\hskip 1pt \vrule width 4pt height 8pt
    depth 1.5pt \hskip 1pt}}
\def \qed{\quad\blackslug\lower 8.5pt\null\par}
\newcounter{mynote}[section]

\newcommand\ignore[1]{}

\newcounter{rcnote}[section]

\newcounter{mrnote}[section]

\newcounter{fknote}[section]

\newcounter{anote}[section]

\DeclareMathSymbol{\mlq}{\mathord}{operators}{``}
\DeclareMathSymbol{\mrq}{\mathord}{operators}{`'}

\newcommand{\rhf}[2]{R_{f, \gamma}}

\DeclareDocumentCommand{\edist}{o o}{
  \ensuremath{
    \IfNoValueTF{#1}{{d}}{{\sf d}(#1,#2)}
  }
}

\newcommand{\olrk}[1]{\ifx\nursymbol#1\else\!\!\mskip4.5mu plus 0.5mu\left(\mskip0.5mu plus0.5mu #1\mskip1.5mu plus0.5mu \right)\fi}

\NewDocumentCommand{\indseq}{ O{1} O{r} }{{#1}\ldots {#2}}

\setcopyright{none}
\renewcommand\footnotetextcopyrightpermission[1]{}

\begin{document}
%\fontfamily{lmr}\selectfont
% \def\thetitle{A Practical Way to Generate Strong Keys from Noisy Data}
\fancyhead{}
\def\thetitle{CTF for education}
\title{\thetitle}

\author{Yi Lyu, Luke Dotson, Nic Draves, Andy Zhang}
\affiliation{\small{University of Wisconsin-Madison}}

\date{}

\begin{abstract}
In this paper, we take a close look at how CTF can be used in cybersecurity education. We divide the CTF competitions into four different categories, which are attack-based CTFs, defense-based CTFs, jeopardy CTFs and gamified and wargames CTFs. 

We start our analysis by summarizing the main characteristics of different CTF types. We then compare them with each other in both learning objectives and other aspects like accessibility. We conclude that combining all four CTF formats can help participants build one’s cybersecurity knowledge.

By doing that, we hope that our findings will provide some useful insights for future CTF educators.

\end{abstract}

\maketitle

% Section I
\section{Introduction}
\label{sec:intro}

%Capture the flag challenges are a great way to learn, practice, and test your knowledge of cybersecurity topics. There are many differentof challenges that cover a large number of topics and fields, including Data security, Software security, Component security, Connection security, and System security\cite{cybersecurity_taught_in_CTF}. The challenges can be made for each type of CTF style, but the way these challenges are formatted might change the way people approach and learn from the challenges. With that in mind, we want to look at the different styles of CTFs, and analyze how they are used to teach skills in Cybersecurity, learn the advantages and disadvantages of each style, and see how we can better utilize CTFs to improve teaching skills in cybersecurity.

Capture the flag (CTF) challenges have been a popular and effective way for individuals to learn and practice cybersecurity skills for over two decades. Originating in the late 1990s as a way for computer security professionals to test their skills and knowledge, CTFs have evolved into a widely-used tool for learning and education in the cybersecurity field. There are many types of challenges that cover a large number of topics and fields, including Data security, Software security, Component security, Connection security, and System security\cite{cybersecurity_taught_in_CTF}.

In this study, we aim to assess the effectiveness of different CTF formats in teaching cybersecurity skills by comparing and evaluating four different CTF formats: Attack, Defense, Jeopardy, and Gamified Challenges or Wargames. Through our analysis, we hope to gain a better understanding of how each CTF format handles the implementation of different subject material, the accessibility of the platforms, and the range of difficulties available to users. Our ultimate goal is to determine if CTFs are a useful learning tool and how each format can benefit individuals interested in cybersecurity education. By examining the strengths and weaknesses of each CTF format, we hope to provide insights into how CTFs can be used more effectively as a learning tool in the cybersecurity field.

% Introduction to your project. Start from some common knowledge that most of the
% reader (in computer security) would have and then narrow down to the details of
% your project. Speak about why the project is important, and why the reader
% should care about it. Finally talk briefly about what are you have done (for
% final project), what you are planning to do (for proposal). Reader should get a
% good chunk of understanding about your project from this introduction section
% (\secref{sec:intro}).
% It's good to finish introduction section with a quick list of contributions. 

% Capture the flag challenges are a great way to learn, practice, and test your knowledge of cybersecurity topics. There are many different kinds of challenges that cover a large number of topics and fields, including Data security, Software security, Component security, Connection security, and System security [1].

% \subsection{For project proposal.}\label{sec:proposal}
% The sections mentioned here is just for reference. You are free to change them
% as you find suitable. In particular for proposal, some of the sections such
% as~\secref{sec:eval} might not make much sense. You can skip that. 

%%% Local Variables:
%%% mode: latex
%%% TeX-master: "main"
%%% End:

%  LocalWords:  biometrics cryptographic parallelized lossy
    % basic introduction
\section{Background and Related Work}
\label{sec:relwork}

In order to give the students more hand-on experiences about cybersecurity, universities are giving the students some cybersecurity exercises\cite{explore_cybersecurity} to practise. How to design high quality cybersecurity exercises is vital to cybersecurity educators.

With the rise of CTF (Capture-the-flag) challenges, many researchers are hoping to find a way to turn CTFs into a useful tool for cybersecurity education. CTFs are an inter-academy competition\cite{cybersecurity_definition} in which each team design and manage a network of computers. By capturing the flag, the participants gain scores and get a higher rank on the leaderboard. CTF challenges are mostly in real time and bring huge excitements to participants.

Victor and Adrian from Military Technical Academy proposes a few feasible guidelines to design cybersecurity exercises\cite{guide_for_desigining_exercises}\cite{efflex}\cite{vetrass}\cite{catp}\cite{monom}. They summarize the steps as defining the objectives, choosing an approach, designing network topology, creating a scenario, establishing a set of rules, choosing appropriate metrics and learning lessons\cite{guide_for_desigining_exercises}. 

Stylianos from Ionian University\cite{storytelling} suggests that we need to present storytelling challenge settings in order to attract the participants. Also the instructors should select suitable CTF platforms and design the exercises according to the preferences and skills of the participants.

Davis et al.\cite{Fun_and_Future} suggested that attack-based CTFs are growing in popularity and will have a larger growth than defense-based CTFs. The reasons they provide include attack only CTFs having smaller scale, participants can have more fun, less player stress, and lower efforts when organizing events.

% In~\secref{sec:intro} you talked about the project at a very high-level. This is
% the section from where you will start giving details. First with things that are
% already done, and familiarize the reader with background information they will
% need to understand you work. 

% Often this section you will discuss the threat-model, but there is no strict consensus on that. 

% Here is how you cite papers. For example, we read papers in the
% class~\cite{rahul2016pwtypos,dodisetal:2004}.  And here is some random citation~\cite{Bojinov:2010:KLP,schechter:2010:pen,everspaugh2015pythia,bellare2009format,Juels:2014}

% \subsection{Overview of the design}
% \label{sec:overview}

% And then just to showoff some \LaTeX skills, here is a Tikz plot.
% \input{images/mainflow.tex}

% You refer to a figure in the following way. In~\figref{fig:mainflow} we show
% some thing that is relevant for the Multisketch paper by Chatterjee et
% al.~\cite{chatterjee2019multisketches}. Add your bibliography to the
% \textsf{bib.bib} file. You can copy the Bibtex format citation from Google
% Scholar.

%%%%%%%%%%%%%%%%%%%%%%%%%%%%%%%%%%%%%%%%%%%%%%%%%%%%%%%%%%%%%%%%%%%%%%%%%%%%%%%%
\section{Methodology}
\label{sec:methodology}

In order to have a systemized approach for gathering information and comparing CTF challenges, we separated the challenges into 4 larger categories: \textbf{Attack}, \textbf{Defense}, \textbf{Jeopardy}, and \textbf{Gamified Challenges \& Wargames}.

For each of these categories we looked into literature as well as relevant websites and competitions in order to collect information to compare them. We wanted to observe how each category handled implementing different subject material (such as cryptography, reverse engineering, or web exploitation), as well as the frequency of each subject. We also wanted to compare the accessibility of the platforms as well as the range of difficulties available to users. Through all of this our final goal was to judge if CTFs are a useful learning tool and how each of these formats could benefit users interested in furthering their cybersecurity education.

% Use this section describe the main contribution of your paper. If you are
% building something, describe the design of the system you built. If you are
% measuring something (like the world!) then include the measurement pipeline.

% This is typically the largest section in your paper. (In case of measurement the
% result section might be bigger.)

% Make sure you give enough details so that the reader is able to reproduce your
% work.

%%%%%%%%%%%%%%%%%%%%%%%%%%%%%%%%%%%%%%%%%%%%%%%%%%%%%%%%%%%%%%%%%%%%%%%%%%%%%%%%
\section{Results}
\label{sec:eval}

\subsection{Attack based CTFs}

For attack based CTFs, the participants act as attackers and gain points by attacking the targets or their opponents. Through this experience, the participants gain a better idea about how to defend against attacks\cite{guide_for_desigining_exercises}. A typical attack workflow can be shown in figure \ref{workflow_attack} below.

\begin{figure}[h]
\includegraphics[width=4cm]{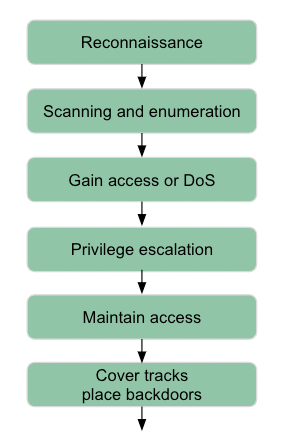}
\caption{Typical attack based CTF workflow \cite{guide_for_desigining_exercises}}
\label{workflow_attack}
\end{figure}

Starting from reconnaissance, the participants determine which kind of vulnerabilities could exist in a certain target. Then by scanning and enumerating different combinations of possible vulnerabilities, the participants can gain more insights about the target and decide which kind of attacks to perform. Following that, they try to get access of the target systems and try to get privilege escalation in order to steal secret information, called flag in CTFs. Finally they submit the flags to earn a score.

A few popular frameworks for attack based CTFs are Bettercap,  Binary Ninja, Metasploit. Bettercap is used to perform man-in-the-middle attacks. Binary Ninja is used for analyzing binary files. And Metasploit is a more general attack based exploitation tool.

There are three popular content categories for attack based CTFs, which are cryptography/hash cracking, web exploitation/network analysis and also reverse engineering/binary exploitation.

The learning objectives of attack based CTFs is to develop the students' abilities to discover vulnerabilities and also apply the knowledge learned to attack given target. It can also improve the students' awareness to write safer codes when deploying their service.

One potential downside of attack based CTF is that it focuses less on defensive side of cybersecurity. In order to build a safe system, you need to think more broadly about all potential risks. But for attacking a system, you just need to find one workable solution.

\subsection{Defense based CTFs}
Defense-based CTF events, also known as cyber defense exercises, are competitions that focus on the perspective of a system administrator. In these events, participants are tasked with securing services and maintaining the functionality of an already-deployed system, while also detecting and responding to outside threats and fixing any bugs or vulnerabilities that they may find in the system or server\cite{Fun_and_Future}. In these CTF events, the red teams are in charge of simulating real-world cyber attacks, and are often played by third parties including volunteers and security professionals.

In the United States, there are three major defense-based CTF competitions: the National Collegiate Cyber Defense Competition (CCDC), CyberPatriot, and Hivestorm. One example of a defense-based CTF event we will look into is CCDC, which challenges student teams to protect and maintain the functionality of a simulated small company's network. Teams are given an identical set of hardware and software and are scored on their ability to detect and respond to outside threats, maintain the availability of existing services, and balance security needs with business needs. An automated scoring engine is used to verify the functionality and availability of each team's services periodically, while a volunteer red team simulates real-world cyber attacks\cite{CCDC}. This allows teams to practice their defensive skills against live opponents and gain valuable experience in protecting against cyber threats.

These CTF games allow participants to practice their system administration skills and learn how to respond to different types of cyber threats. As cyber warfare has become an increasingly important focus for military and government organizations in recent years, defense-based CTF events provide a valuable opportunity for professionals to gain hands-on experience in defending against cyber attacks.

Defense-based CTF events can be a valuable learning experience for professionals looking to gain experience in cyber defense and enhance their system administration skills. However, there is one large limitation to this type of CTF. The potential shortcoming is that defense-based CTFs are often organized in the form of competitions, which can require significant time and resources to set up the relevant servers and clients. This can make them less accessible than other types of CTFs, which may be easier to set up and participate in.

\subsection{Jeopardy CTFs}
Jeopardy CTF’s are a type of CTF based on the TV show Jeopardy, where cybersecurity questions are split into categories based on subject material (Cryptography, Binary Exploitation, Web Exploitation), and participants are awarded points based on successful completion of the tasks given to them. Jeopardy CTF’s are frequently used as the first round for live competitions \cite{framework}, such as Defcon CTF and PHD CTF, as a way to establish that teams moving to later rounds have a solid basis in cybersecurity knowledge needed to move on to later rounds of the competition.

While the Jeopardy CTF format obviously has it’s benefits for competitions, it is perhaps more pertinent to the topic of CTFs as a learning tool to look into the websites that host Jeopardy CTFs available at all times that students could work on. For this paper, the two examples of online Jeopardy CTF’s that we will look into are picoCTF and 24/7 CTF.

PicoCTF (found at picoctf.org) is a website hosted by Carnegie Mellon University which hosts a yearly Jeopardy style CTF competition online. On top of this competition, picoCTF has a large pool of practice questions that users can complete to earn points in the section of their website labeled “picoGym.” The picoGym offers over 270 practice exercises rewarding a variable number of points, (between 10 and 500)  based on difficulty for the completion of the exercise. Pico also provides a web shell so that users do not have to use a virtual machine to run challenges, helping with user accessibility. Of the challenges hosted on picoCTF, 17\% are web exploitation challenges, 18.5\% are cryptography challenges, 20\% are reverse engineering challenges, 18.5\% are forensics challenges, 13.5\% are binary exploitation challenges and the remaining 22.5\% are in the general skills or uncategorized categories \cite{pico}.

24/7 CTF (found at 247ctf.com) is a community run Jeopardy style CTF that contains an online leaderboard to keep track of global users scores. 24/7 CTF hosts 72 total challenges including an 8 challenge beginner tutorial to help users get familiar with capture the flag challenges. While the beginner tutorial is certainly a point in favor of the site’s accessibility, it does run into some other issues, namely that some challenges require the use of a VM as it does not provide a web shell, and that hints are hidden behind a paywall. Of the non tutorial challenges on the site, 14\% are cryptography, 12.5\% are networking, 21\% are web exploitation, 14\% are reverse engineering, 20\% are labeled “pwnable”, and the remaining 18.5\% are miscellaneous or uncategorized \cite{247}.

While Jeopardy CTFs overall provide a solid foundation of skills for participants, particularly in introducing students to new categories, they do have some downsides as a learning tool. The most obvious drawback of Jeopardy style CTFs is that since the questions are very compartmentalized, they do not give users the experience of looking for exploitations or patching vulnerabilities in a real environment in the way that other CTF’s do. 

\subsection{Gamified Challenges or Wargames}

Compared to the other categories of CTFs covered in this paper, Gamified challenges and Wargames have a few notable differences that make them a unique option for learning Cybersecurity. In this paper, we will be looking at both of these styles together since even when combined, they are still less popular than all the other categories that have been previously brought up in the paper. Although there are a smaller number of people learning from gamified and wargame style challenges, there are still a good number of challenges and options to learn in these styles, while also offering features and formats often not seen in other styles of CTFs.

Starting with some similarities, both of categories have don’t have a time limit or a time factor for challenges. Instead, challenges in this style are built so that the player can take as much or as little time as they need to complete the challenge. Having no time limit encourages users to explore the systems they are attacking and allows for a greater amount of time to focus on information gathering. Another similarity between both Gamified challenges and Wargames is that they often cover any topic, not just common categories or well known exploits. This means that these systems could be build for general challenges such as SQL injection, or with very specific challenges such as identifying a system with an uncommon vulnerability and then requiring the user to exploit the vulnerability to continue. Due to some challenges having very focused topics, they can be a bit difficult for new users, but multiple difficulties of each challenge styles are available in both gamified challenges and wargames. 

Gamified Challenges approach topics with a focus on story driven teaching or with a themed set of challenges. With these challenges, the player works on topics while progressing though increasingly difficult challenges and tasks until either the system is fully comprised or the end of the story in the challenge box is reached. These types of challenges are common from services such as Vulnhub, which has a large number of user contributed gamified VM challenges\cite{10.1007/978-3-030-63396-7_34}. The VMs on this platform are commonly themed or very topic focused and have the user use whatever Cybersecurity skills are needed to complete the challenge. The VM gamified challenges are a fun and unique way to learn and improve upon hacking skills using real tools and exploits. However, other gamified security and privacy challenges exist that don’t go very far in depth, and instead focus on introducing new topics, cover attack vectors, teach threat modeling, and other privacy and security topics\cite{LearningGamification}. These gamified challenges are often gamified security training, video games integrating cybersecurity topics, and themed card \& tabletop games.

Wargames take a different approach to Gamified challenges in a few different ways. Most wargames require the user to connect to external labs to start challenges. This means that instead of downloading a VM or connecting to a website in a web browser, the user is instead required to connect and work on challenges exclusively in the terminal. Once connected, a player would either have access to one machine with multiple challenges to solve, or a lab with many systems. In a lab, the player would set out to either compromise all of the computers, complete challenges that require the talking to other systems, or solve networking based challenges. With wargames being built with the focus on connecting to real systems, the skills being learned in this style have a very heavy focus on realism, and are great tools to understand how an attacker would go about exploit software and hardware bugs in a small network \cite{smashthestack}. Another unique part of wargames is that a lab might have multiple users in it, which means that  might not be the only attacker trying to complete challenges on the network. Wargame platforms, such as Over The Wire and Smash the Stack, offer a very unique learning experience that often can often only be learned by setting up a personal lab.

\subsection{Comparison of different CTF platforms}

For students who are new to CTF challenges, they often lack the experiences about setting up the challenge environment, finding the hints in the question settings and also flag submission process. Thus it is important to choose a platform which has good documentation and also easy to set up the environment. Based on the result of Tanner and his team\cite{CTF_analysis}, we compare the platforms from three different perspectives, how to install, what the language the students are going to use, and also how well the documentation is.

\begin{table}[h]
    \begin{center}
        \begin{tabular}{|c|c|c|c|} \hline
             & Installation & Language & Documentation \\ \hline
            PicoCTF & Vagrant$\checkmark$ & Python$\checkmark$ & Good$\checkmark$\\ \hline
            OpenCTF & Docker$\checkmark$ & Python$\checkmark$ & Simple\\ \hline
            CTFd & Native & Python$\checkmark$ & Simple\\ \hline
            FbCTF & Vagrant$\checkmark$ & PHP & Good$\checkmark$\\ \hline
            TinyCTF & Native & Python$\checkmark$ & Simple\\ \hline
        \end{tabular}
        \vspace{0.5cm}
        \caption{Comparison between different CTF platforms' accessibility\cite{CTF_analysis}}
    \end{center}
    \label{tab:platform_accessibility}
\end{table}

Among five CTF platforms, there are three different installation methods, Vagrant, Docker and Native. For beginners, installing on native machine may be challenging since there are many things that can go wrong. Everyone can have different environment settings and it may require extra efforts to debug the installation when things go wrong. Also it is not safe to install vulnerable services on the students' machine. A better approach is to use virtualized environment. Both Vangrant\cite{vangrant} and docker\cite{docker} are an easy way to create and configure lightweight, reproducible, and portable development environments. The students do not need to worry about configurations and can set up a uniform environment easily.

Then we compare differnet languages used in CTF challenges in these platforms. There are two main languages, Python and PHP. Considering that Python has a higher popularity and usage in both the academia and industry these days, it would be easier for the students to use Python.

Finally let's compare the documentations of the platforms. They considered both the installation documentation and challenge documentation. Some platforms give relatively simple documentation compared with other platforms.

Based on the findings above, we believe that PicoCTF has the best accessibility among all four other platforms and is most suitable for teaching.

\subsection{Comparisons between different CTF types}
One of the most valuable ways to provide insights into the benefits of each CTF format is to draw direct comparisons between them to focus on their strengths and weaknesses. These formats can vary in terms of user accessibility, applicability to real life situations, and the perspective of the participants in relation to vulnerabilities.

For ease of direct comparison, we will separate this section into Attack based vs. Defense based CTFs and Jeopardy CTFs vs Gamified Challenges \& Wargames. This is because for Attack and Defense based CTFs, the strengths and weaknesses are largely due to the perspective of the participants, either as someone who is looking to exploit a vulnerability or a system administrator looking to prevent an exploit or patch a vulnerability. This is different from the benefits and deficiencies of Jeopardy CTFs and Gamified Challenges, as the comparisons between those categories will largely be in terms of accessibility to participants, depth of topics covered, and the applicability to real life scenarios.

\subsubsection{Comparison between Attack and Defense based CTFs}

For attack based CTFs, the participants work from an attacker's and find out the vulnerabilities to attack the targets. The common strategies to find out the vulnerabilities is scanning and enumeration. The participants need to design probing strategies, observe the feedback of the target and determine if there is a vulnerability in this part. This requires the participants to have a certain domain knowledge of the system they are going to attack. When the search space is huge, they also need to design efficient approaches to narrow down the search space.

Similarly, the system administrators on the defensive side also need to find vulnerabilities or bugs and fix them before the red team exploits them. Although different kinds of vulnerabilities have different levels of difficulty to be exploited, they would show the same level of threat the time they are cracked. Thus, in addition to fixing bugs as fast as possible, the participants also need to block every possible kind of entry to ensure that the system is secure.

On the scope of cybersecurity education, the two CTFs have different focuses. While attack-based CTFs focus more on the efficiency of the attacks, defense-based CTFs focus on the completeness of protection. With different focuses, the participants can have different mindsets when participating in the two kinds of events. Since both the skills and mindsets of the two CTFs are complementary, participants can have a better understanding of cybersecurity when taking part in both kinds of events.

\subsubsection{Comparison between Jeopardy and Gamified challenges \& wargames}
When comparing Jeopardy style CTFs to Gamified challenges and Wargames, the areas where the two differ the most are in their depth, accessibility, and contextualization to real systems. In Jeopardy CTF's, the challenges tend to be relatively short in length, regardless of difficulty. This is likely in order to give users the chance to complete more challenges, either online or in a live competition, as Jeopardy is meant to test a breadth of knowledge. On the contrary, many Gamified challenges, particularly Wargames, tend to have the user find specific vulnerabilities that build on each other in order to have a deeper understanding of the topic that they are covering.

Another key difference between Jeopardy and Gamified CTFs is in the accessibility to users. With many Jeopardy CTFs, the barrier to entry can be as low as entering the web page, as it requires little to no downloading of additional materials and doesn't necessarily need a virtual machine to run. On the other hand, due to the more realistic nature of Wargames, many require you to have a virtual machine to run the material, and can make several assumptions on technical skills of the user not directly related to the material being covered.

The final point of comparison between Jeopardy CTFs and Gamified CTFs is in terms of contextualizing the vulnerabilities to larger systems. Jeopardy CTFs will normally provide a problem without any context to the system that it is found in, which can gloss over several of the challenges of working on cybersecurity in live environments. In comparison, Wargames often will put you directly into a simulated system, leaving it up to you to search out and find vulnerabilities in a way that provides context much closer to what would be found working on a real system.

% The results section should have the results of your experiments and
% measurements. Evaluation is the most important part of the research. Sometime
% you might want to split it into more than one section depending on the type of the project.

% Again make sure you give enough details so that the reader can reproduce your
% evaluation. Your GitHub code is not a replacement of the details, GitHub will
% perish, your paper will remain in this world for centuries to come. Of course,
% you need strike a balance between mundane details vs what makes your evaluation
% unique.

\section{Conclusion}
\label{sec:conclusion}
With nearly half of UW Madison Computer Science students not taking a course in security before they graduate, we know firsthand that many developers in the workforce are not sufficiently prepared to deal with cybersecurity issues in the real world. CTF challenges provide a learning tool that allows anyone with sufficient computer science experience to have a safe environment to learn about exploiting vulnerabilities and defending against attackers without needing oversight from a third party. CTF challenges also work as a great supplement to existing cybersecurity experience by allowing users to explore topics that they are less familiar with, and can also guide users to learning more advanced techniques in a field where they already have experience.

Throughout the course of this paper, we have talked about the strengths and limitations of each of the categories of capture the flag challenges. While we acknowledge the limitations of these challenges, we believe that many of these limitations can be overcome by working on several different kinds of CTFs, as many times the weakness of one type of CTF is the strength of another.

While no single CTF challenge or platform can provide everything a user can need to know about cybersecurity, by participating in a range of CTFs users can build a holistic basis of cybersecurity knowledge, and are given a tool for continuous improvement of their security skills.

% What is the big take away from your research. Include any limitations or future work here. 

There are still some limitations regarding our research in this project. Our paper does not provide much information on the effectiveness of CTFs as a teaching tool in a quantitative way. Also, the comparison between different CTF formats relies primarily on literature and websites and does not include much user feedback.

Thus, future researchers can address some of these limitations by analyzing the effectiveness of different CTF formats for teaching cybersecurity skills quantitatively, studying how different CTFs can be used as teaching tools for users with different levels of experience, or developing new and innovative CTF formats to improve user engagement.

%%% Local Variables:
%%% mode: latex
%%% TeX-master: "main"
%%% End:

% \section{Contribution}
% \label{sec:contribution}
% Luke Dotson took the lead on researching Jeopardy CTFs, as well as taking part in searching for relevant literature for the team. He also led the  methodology and conclusion sections of the paper and contributed to the comparisons between Jeopardy and Gamified CTFs.

% Andy Zhang focused his research on defense-based CTFs and searched for some papers on the topic of CTF as a whole. He also contributed to the introduction and comparisons between attack and defense-based CTFs.

% Yi Lyu did her most research on attack-based CTFs, as well as taking part in searching for relevant literature for the team. She also wrote the background and related work part. In addition, she contributed to the comparison between different CTF platforms and  the comparisons between attack and defense-based CTFs.

% Nicolas Draves did the research for the gamified challenges and wargames and finding relevant papers that focused on these topics specifically. Additionally, he helped planning and writing comparisons for “comparisons between Jeopardy and Gamified challenges”

\bibliographystyle{ACM-Reference-Format}
\bibliography{bib}

% % --- Appendix ---%
\appendix
% \section{Overflow form other sections}
% \label{sec:set-diff-dodis}
% Sometime you ware super excited about some details that does not quite fit with
% the rest of the paper goes here. For example, some details about how you
% instrumented the Android Linux kernel should go to appendix, and for really
% curious reader to read. Remember it's appendix, so the reader is not required to
% read, and you should not put critical information in appendix that is crucial
% for understanding the rest of the paper.

%%% Local Variables:
%%% mode: latex
%%% TeX-master: "main"
%%% End:

\end{document}